\documentclass[ aps,pre,twocolumn,showpacs,superscriptaddress]{revtex4-1}  % for review and submission
\usepackage{graphicx}  % needed for figures
\usepackage{dcolumn}   % needed for some tables
\usepackage{bm}        % for math
\usepackage{amssymb}   % for math
\usepackage{amsmath}
\usepackage{epsfig}
\usepackage[utf8]{inputenc}

\usepackage{color}

\begin{document}

\title{
 Universal avalanche statistics and triggering close to failure in a mean field model of rheological fracture
}

\author{Jordi Bar\'o}
\email{jordi.barourbea@ucalgary.ca}
\affiliation{Department of Physics and Astronomy
University of Calgary.
2500 University Drive NW
Calgary, Alberta  T2N 1N4, Canada}

\author{J\"orn Davidsen}%
\affiliation{Department of Physics and Astronomy
University of Calgary.
2500 University Drive NW
Calgary, Alberta  T2N 1N4, Canada}

\begin{abstract}
The hypothesis of critical failure relates the presence of an ultimate stability point in the structural constitutive equation of materials to a divergence of characteristic scales in the microscopic dynamics responsible for deformation. Avalanche models involving critical failure have determined common universality classes for stick-slip processes and fracture.
However, not all empirical failure processes exhibit the trademarks of criticality. 
The rheological properties of materials introduce dissipation, usually reproduced in conceptual models as a hardening of the coarse grained elements of the system. Here, we investigate the effects of transient hardening on (i) the activity rate and (ii) the statistical properties of avalanches. We find the explicit representation of transient hardening in the presence of generalized viscoelasticity and solve the corresponding mean field model of fracture. In the quasistatic limit, the accelerated energy release is invariant with respect to rheology and the avalanche propagation can be reinterpreted in terms of a stochastic counting process. A single universality class can be defined from such analogy, and all statistical properties depend only on the distance to criticality. We also prove that inter-event correlations emerge  due to the hardening --- even in the quasistatic limit --- that can be interpreted as ``aftershocks'' and ``foreshocks''. 
\\

\end{abstract}


\maketitle{}

\section{Introduction}

The mechanical failure of natural or man-made structures due to the variations of the external loads or long exposure to extreme external conditions constitutes a common hazard of major concern in seismology and civil engineering. Experimental studies reveal that the mechanical deformation of crystalline structures \cite{Li2002,Friedman2012}, amorphous materials \cite{Antonaglia2014,Salerno2015} and jammed granular (or fragile) matter \cite{Hidalgo2002,Denisov2017} is highly affected by the inherent heterogeneity in the system \cite{Bonamy2011} or some degree of disorder such as defects, dislocations or inclusions. Hence, failure is difficult to forecast because of the sensitivity to the unknown internal details of the system. 
In micromechanical models of failure \cite{Mogi1962,Burridge1967,Olami1992,Alava2006,Benzion2011,Girard2010,Shekhawat2013,DeArcangelis1985,Duxbury1987,Zapperi1997,Csikor2007,Amitrano2012,Moreno2000,Girard2010,Kun2000}, the addition of disorder is able to arrest the internal micromechanical processes responsible of deformation in multiple metastable states, leading to stochastic avalanche dynamics \cite{Sethna2001}. The arrested energy is partially released during the mechanical avalanche as elastic waves that can be detected by means of seismographs and geophones at the geological scale~\cite{Benzion2008,Kanamori2004}, or by ultrasonic acoustic emission (AE) equipment \cite{Scruby1987,Lockner1993} in laboratory controlled experiments (see, for example, \cite{Main1989,Hidalgo2002,Scholz1968,Meredith1983,Lockner1993,Anifrani1995,Davidsen2007,Michlmayr2012,Goebel2013,Baro2013,Ribeiro2015,Costa2016,Davidsen2017}). Such elastic waves can be used as probes to assess the state of the system and develop reliable forecasting tools for structural health monitoring \cite{Grosse2008,Grosse2003}. Mechanical failure appears as a consequence of weakening or yielding of the strain ($\varepsilon_{ij}$) stress  ($\sigma_{ij}$) relation. The susceptibility of the strain to variations of the stress tensor --- a parameter simplified in this work by a scalar modulus $G=d\varepsilon/d\sigma$ --- increases as the materials weakens. As a consequence, the energy released as elastic waves --- trademark of the amount of deformation --- also increases close to failure. This increased energy release has been claimed to be present in the vicinity of major earthquakes --- or mainshocks --- and it is usually referred to as accelerated moment or seismic  release (AMR or ASR) in both seismology and AE experiments \cite{Anifrani1995,Jaume1999,Benzion2002,Zhang2006,Wang2008}, although its validity in seismology is controversial~\cite{Vere-Jones2001, Benzion2002,Mignan2011}, and rarely outscores linear models as a forecasting tool \cite{Robinson2005}. 
Several micromechanical models governed by quenched disorder justify the observation of ASR by the presence of a critical point matching failure \cite{Sornette1995,Jaume1999,Kun2000,Amitrano2012,Amitrano2005,Alava2006,Dahmen2011,Vasseur2015}. Again, in the case of seismology, this hypothesis might be questionable~\cite{Sornette1995,Hardebeck2008,Ramos2010}. In the presence of a critical point, close to criticality the distribution of avalanche energies ($E$) can be described by a generalized homogeneous function: 
\begin{equation}
D(E;f)dE = E^{-\epsilon} \mathcal{D}(E f ^{\beta}) dE 
= f^{\beta \epsilon}\tilde{\mathcal{D}}(Ef^{\beta})
dE.
%D(E;f)dE \sim E^{-\epsilon} e^{-A(E f ^{\beta})} dE = f^\frac{\beta %\epsilon}{\gamma}\tilde{\Phi}(Ef^\frac{\beta}{\gamma})
%dE
\label{eq:BD:critFail}
\end{equation}
with a scaling function $\mathcal{D}$ that depends only on the combined argument  $Ef^{\beta}$. Here, $f$ accounts for the distance to failure and is defined in terms of the time remaining to reach the failure point $f=1-t/t_{f}$. In critical failure models, the observed ASR is a natural consequence of the increase of the mean event energy: If the activity rate is constant, the energy rate $\frac{d E (f)}{df}$ will be proportional to $\langle E(f) \rangle \sim f^{{(\varepsilon-2)}{\beta}}\sim (t-t_{f})^{{(\varepsilon-2)}{\beta}}$. 
Although widely accepted \cite{Mignan2011,Amitrano2012} this explanation is insufficient in the presence of non-conservative processes, which are known to play an important role in rock fracture~\cite{Bonamy2011}. The addition of dissipation introduce length-scales and can prevent criticality \cite{Vespignani1998}, as specifically shown in branching processes~\cite{Lauritsen1996}, stick-slip models~\cite{Mehta2006} and depinning interface models~\cite{Jagla2010}. \\

Instead of, or in addition to, criticality, almost all experimental studies show an increase in the number of events coinciding with failure~\cite{Main1989,Liakopoulou1994,Lennartz2014,Davidsen2017} and/or large events~\cite{Baro2013,Ribeiro2015,Costa2016}, typically denoted as the inverse Omori law~\cite{Utsu1995,Ojala2004}. Such a behavior at the failure point is not reproduced by standard micromechanical models and the provided analytical solutions implicitly consider constant activity rates~\cite{Kloster1997,Benzion2011}. As shown experimentally~\cite{Mogi1967,Benzion2002}, ASR can simply be a consequence of this increase in the number of events alone.  \\

In the present manuscript we argue that the same processes responsible for the observed history-dependent activity~\cite{Lockner1993,Davidsen2017,Baro2013,Ribeiro2015,Costa2016}, namely relaxation mechanisms \cite{Dieterich1994,Benzion2006,Burridge1967,Hainzl1999,Nakanishi1992,Shcherbakov2004,Benzion2006, Mehta2006,Kawada2006,Jagla2010,Jagla2014,Lippiello2015} often related to event-event triggering \cite{Ogata1988,Gu2013,Zaliapin2013,Baro2017}, can explain ASR as peaks of activity, even in absence of critical failure or any temporal variation in the statistical properties of the AE events. We show that the emergence of relaxation processes and the associated temporal correlations can be a direct consequence of dissipation as modeled by transient hardening. We mathematically explain the link between aftershocks, foreshocks, critical failure and accelerated seismic release at a fundamental level in a solvable model of fracture.
\\

The starting point of our study is a prototypical model of fracture: the democratic fiber bundle model (section \ref{sec:BD:stdDFBM}).
Incorporating experimental findings, we propose a variation of the model with a physically based transient effect, which we denote as the generalized viscoelastic democratic fiber bundle model (section \ref{sec:BD:veDFBM}) able to generate
 relaxation processes and triggering (section \ref{sec:BD:as}).
We prove analytically that this model can be approximated to the more simple and general concept of transient hardening (section \ref{sec:BD:trans}).
We derive the mean field (MF) solution of the transient hardening model in the thermodynamic limit (section \ref{sec:BD:MF}). In the process, we reinterpret the model as a fundamental stochastic problem. We find that a unified universality class (UC) for fracture models can be derived from this model, which is distinct in its initial formulation from the MF model of slip avalanches \cite{Benzion2011} and critical branching processes \cite{Zapperi1995}. 
 In the presence of transient hardening, the critical point is never reached. The magnitude of transient hardening and the distance to the failure point are combined in a single parameter --- the distance to criticality--- that fully determines the characteristic scales of avalanche statistics. 
We test our analytical results with numerical simulations of the viscoelastic model.
The numerical findings for the standard viscoelastic case are presented in section \ref{sec:BD:num}.
The temporal evolution of the distance to criticality during the failure process depends on the driving mechanism (section \ref{sec:BD:B}).
  The function of the stochastic sampling and the magnitude of the hardening completely define the avalanche size distribution (section \ref{sec:BD:distro}), the evolution of the activity rate and the seismic release of the process leading to failure (section \ref{sec:BD:foreshocks}). 
  We observe Omori-like behavior --- typically associated with triggering and aftershocks~\cite{Ogata1988,Utsu1995} --- with self-consistent specific exponents (section \ref{sec:BD:trigger}).  
 We comment on the implications for the experimental observations and present some concluding remarks in section \ref{sec:BD:conclusions}.

\section{Methods}
\subsection{The standard democratic fiber bundle model (DFBM)}
\label{sec:BD:stdDFBM}

\begin{figure}
    \includegraphics[width=1.0\columnwidth]{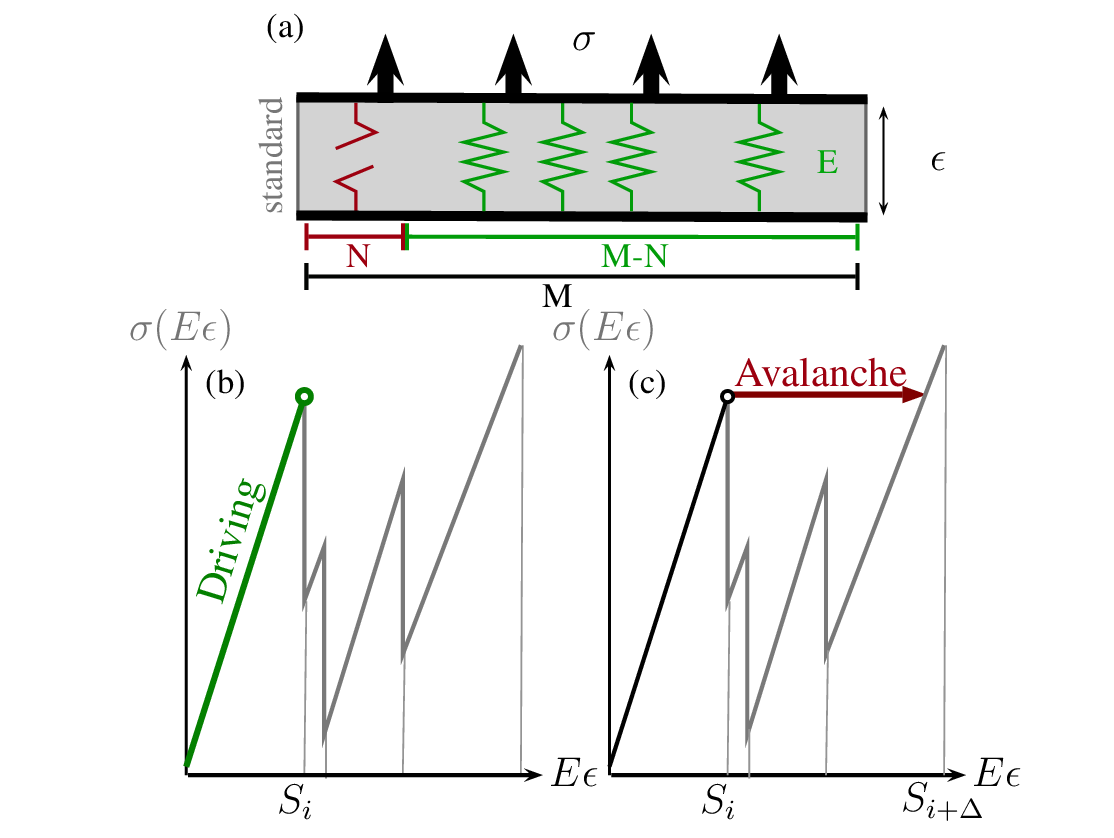}\\ 
\caption{\label{fig:BD:std} (a) Schematic representation of the standard DFBM constituted by an ensemble of $M$ parallel elastic-brittle elements, $N$ of which are broken because their $S_i< E \varepsilon$; (b,c) Sketch of the strain release $E \varepsilon$ due to the breaking of a fiber $S_{i}$ under quasistatic driving, generating an avalanche  of size  $\Delta$. (c) The avalanche stops when the system regains stability, as represented by the constitutive curve (gray line).}
\end{figure}

The democratic fiber bundle model (DFBM) is arguably the simplest model able to reproduce avalanche statistics in irreversible fracture mechanics \cite{Kloster1997,Pradhan2002,Moreno2000}. As represented in Fig.~\ref{fig:BD:std}.a, 
fiber bundle models simulate the mechanical response ($\varepsilon(t)$) to a tensile stress ($\sigma(t)$) of a bundle of $M$ initial fibers ($l$) sharing an externally controlled load. Each fiber is modeled as a coarse grained elastic element with an equal Young's modulus ($E$) and an independent random limit tolerance to deformation, or strength, $S_{i}$ usually sampled from an extreme value Weibull distribution with cumulative distribution: $F(S_{i}<s) = \int_{-\infty}^{s} p(s')d s' = 1 - \exp(-s^{m})$, where $p(s')$ is the probability density function. 
Hence, the local stress: $\sigma_{l}(t) \equiv M x_{l}(t) \times  \sigma(t)=E\varepsilon(t)$, where $x_{l}(t)$ is the fraction of the external stress sustained by the element $l$, and $\sum_{l} x_{l}(t)=1$. Each fiber will break when $E\varepsilon_{i}(t) \geq S_{i}$, setting its contribution to the load $x_{i}(t)\to 0$ and effectively increasing the average $x_{l}(t)$ for the rest of the ensemble. The non-linear response of the bundle emerges from the coupling between the values of $x_{l}$ due to the  brittleness $S_{i}$ of the individual fibers. A good general review on fibrous models can be found in Ref.~\onlinecite{Kun2006}.  
The democratic fiber bundle model (DFBM) corresponds to the mean field solution where all intact fibers contribute equally to the load. The contribution of each fiber can be expressed as a function of the number $N(t)$  of failed fibers over time such that: $x_{l}(t)=\frac{1}{M-N(t)}$ for all $l$. Since all fibers have the same local load, the number of failed fibers 
%due to a monotonically increasing stress (given a maxiumal stress value) 
at a given strain value will be given by the number of fibers with strength $S_{i}<\sigma_{l}(t)$ and thus: $E\varepsilon(t)= \frac{M}{M-N(S_{i}<E\varepsilon(t))} \sigma(t)$. Using the numerical cumulative distribution $F(S_{i}<E\varepsilon)$, the constitutive equation describing the mechanically stable solutions of the DFBM reads:
\begin{equation}
    \sigma(E\varepsilon) = \left({1-F(S_{i}<E\varepsilon)}\right) E\varepsilon{} .
    \label{eq:BD:adiabatic}
\end{equation}

Mechanical avalanches will occur as consequence of the metastable solutions in \eqref{eq:BD:adiabatic} introduced by the profile of $F(S_{i}<E\varepsilon)$ under certain driving conditions. Since $F(S_{i}<E\varepsilon)$ is modified at the breaking of one or several fibers, the mechanical avalanche is caused by the brittle failure of the individual fibers. From now on, we will consider avalanches as the collective instantaneous failure of $\Delta$ fibers, being $\Delta$ defined as the size of the avalanche.\\

As a particular case, a macroscopic brittle event will always occur above a stability limit ($\varepsilon_{f},\sigma_{f}$), that we associate with the macroscopic failure point. In the thermodynamic limit, the failure strain value $\varepsilon_{f}$ under stress driving is given by : $ E\varepsilon_{f}:=\frac{1-F(E\varepsilon_{f})}{p(E\varepsilon_{f})}$, where now $F(E\varepsilon_{f})$ is a continuous function defined by the sampling strength distribution. In addition, at the microscopic level, $F(S_{i}<E\varepsilon)$  is a stochastic step-like function and  will introduce a step-wise drop  in \eqref{eq:BD:adiabatic} at the strength $S_{i}$ of each fiber (see gray lines in Fig.~\ref{fig:BD:std}.b,c), giving rise to an avalanche of size $\Delta$. The probability of size $\Delta$ for the DFBM can be obtained as a particular case of the procedure exposed in detail in section~\ref{sec:BD:MF}.\\

Although fiber bundle models were originally designed to simulate the response of fibrous composite materials to tensile stress, successful adaptations towards continuous damage models reported a good agreement with the behavior of shear processes involving plasticity \cite{Hidalgo2001}, stick-slip dynamics \cite{Halasz2009} and even granular materials under compression \cite{Hidalgo2002}. Thus, one can consider the DFBM as a reliable prototypical and solvable mean field model of brittle failure \cite{Kun2006}, able to explain yielding and critical scaling to failure \cite{Kloster1997,Moreno2000}. Under stress driving, the statistics of the DFBM are compatible with \eqref{eq:BD:critFail} as will be discussed in section \ref{sec:BD:MF} in more detail.

\subsection{The generalized viscoelastic DFBM}
\label{sec:BD:veDFBM}

As in most conceptual and numerical failure models, the interactions between elements in the DFBM propagate much faster than the variations of the external conditions --- corresponding to the quasistatic driving limit --- and any other temporal scale of the system. Thus, the transition between stable solutions is driven exclusively by the avalanche dynamics. Since the strength values $S_{i}$ are independent, the avalanches, defined from the instabilities of Eq.~\eqref{eq:BD:adiabatic}, are uncorrelated. 
The temporal clustering observed in nature and experiments can be reproduced in conceptual models by the introduction of a temporal scale interfering with the avalanche propagation. 
For example, correlations have been observed in stick-slip models with dissipation \cite{Hergarten2002,Pradhan2002}, yet they were shown not to be a consequence of event-event triggering but a consequence of slow temporal variations in the Poisson intensity or synchronization \cite{Hergarten2011}. Power-law waiting times can also be artificially constructed by a non-quasistatic driving and a thresholding of the activity \cite{Paczuski2005,Deluca2015,Janicevic2016}, without requiring the involvement of any triggering or aftershock process.\\

Event-event triggering or aftershock sequences and the associated temporal correlations are commonly reproduced by introducing additional temporal scales affecting the propagation of the avalanches, without requiring to break the quasistatic condition. In the case of fracture and stick-slip processes, it has been proposed that temporal scales are introduced by the non-linear rheological or tribological behaviors of the coarse-grained elements of the material \cite{Mehta2006,Jagla2010}. As examples, micromechanical models reproduce aftershock sequences by incorporating rate and state-dependent friction \cite{Dieterich1994}, damage rheology \cite{Benzion2006}, viscoelasticity \cite{Burridge1967,Hainzl1999,Shcherbakov2004,Lippiello2015} or a viscous drag \cite{Nakanishi1992}.
In general terms, a partial delay in the response of the material such as velocity hardening or viscoelastic creep will introduce an effective transient hardening of the thresholds \cite{Benioff1951, Benzion2006, Mehta2006,Kawada2006,Jagla2014} splitting the otherwise instant transition in a cascade of smaller avalanches \cite{Mehta2006}. 
The relaxation of this hardening towards the equilibrium state can give rise to the temporal correlations between avalanches \cite{Smirnov2004}  mimicking those observed in aftershock sequences. Thus, this process can capture the temporal features associated with event-event triggering observed in seismic catalogs \cite{Ogata1988,Utsu1995,Gu2013,Davidsen2016} and AE experiments \cite{Hirata1987,Lockner1993,Baro2013,Nataf2014,Ribeiro2015,Davidsen2017,Costa2016}. On the other hand, this transient hardening corresponds to a dissipation mechanism coupled to the dynamics, thus, affecting criticality. Hence, both the presence of correlations and the lack of criticality at failure might be reproduced by the introduction of transient hardening in micromechanical models that would normally reproduce critical failure.\\

Here, we derive the mean field solutions to a transient hardening model \cite{Kawada2006} by explicitly incorporating generalized viscoelasticity into the DFBM. 
We compare the analytical solutions with the standard DFBM --- i.e. without viscoelasticity --- to understand how this mechanism of transient hardening affects the statistical properties of avalanches. Specifically, we discuss its ability to explain (i) the presence of ASR without a divergence of scales at failure --- i.e. critical failure --- and (ii) temporal correlations between events. \\

The novel aspect of this model is the substitution of the elastic fibers with generalized Zener solid elements \cite{Glockle1994}. These elements are equivalent to a fractal viscoelastic model \cite{Heymans1994}, stable under stress and strain driving and able to describe realistic memory relaxation processes, as observed, for example, in amorphous solids \cite{Hellinckx1994, Schiessel1995}. This viscoelastic model introduces a physically based mechanism of transient hardening (as will be discussed in section \ref{sec:BD:trans}) in the microscopic elements with brittle failure.\\

\begin{figure}
    \includegraphics[width=\columnwidth]{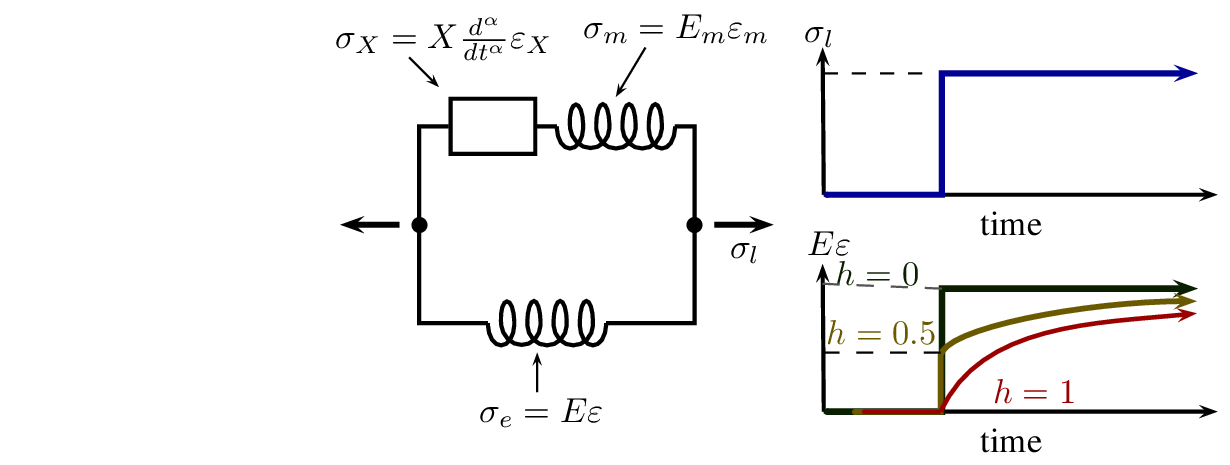}
    \caption{\label{fig:BD:VE} Left: schematic representation of the generalized Zener element. Right: Temporal response $E\varepsilon$ of the generalized Zener element to a stepwise increase in the load $\sigma_{l}$.}
\end{figure}
Each of the elastic brittle elements of the standard DFBM ($\sigma_{e} = \varepsilon E$) is now coupled in parallel to a generalized Maxwell element \cite{Friedrich1991} (see Fig.~\ref{fig:BD:VE}). The generalized Maxwell element consists of a secondary elastic spring ($\sigma_{m}=E_{m}\varepsilon$) coupled in series to a Scott-Blair springpot \cite{Scott1947,Jaishankar2012}, instead of the usual viscous dashpot. While the mechanical response of the standard dashpot reads $\sigma_{X}= \eta d/dt \varepsilon_{X}$, where $\eta$ is the viscosity, the generalized element involves fractional derivatives ($\sigma_{X}= X d^{\alpha}/dt^{\alpha} \varepsilon_{X}$) with physical fractional dimensions $0\le \alpha \le 1$ and a general complex modulus $X$ instead of $\eta$. The constitutive relations of each generalized Zener solid element can be obtained from the mechanical equilibrium between the individual parts as represented in Fig.~\ref{fig:BD:VE}. The conditions $\sigma_{m} \equiv \sigma_{X}$, $\varepsilon \equiv \varepsilon_{m}+\varepsilon_{X}$ and $\sigma_{l} \equiv \sigma_{e}+\sigma_{m}$ have to be satisfied, leading to the constitutive equation:
\begin{equation}
        \left[{1+ \frac{X}{E_{m}}\frac{d^{\alpha}}{dt^{\alpha}}}\right]\sigma_{l}  = \left[{1+\frac{X(E_{m}+E)}{E_{m}E} \frac{d^{\alpha}}{dt^{\alpha}}}\right] E\varepsilon .
\end{equation}
%. Fractional derivatives have history dependent solutions that can be expressed in terms of integrals. Here we will use the Riemann-Liouville formulation:
%\begin{equation}
%    \mathbf{D}^{\alpha}_{t_{0}}f(t)=\frac{1}{\Gamma (n-\alpha)}\frac{d^{n}}{dt^{n}}\left[{\int_{t_{0}}^{t}(t-\tau)^{n-\alpha-1}f(\tau)d\tau}\right]
%\end{equation}
By defining $\tau^{\alpha} :=X (E_{m}+E)/(E_{m}E)$,  
the strain response to a sudden increase in stress $\Delta \sigma$ a time 0 --- i.e. creep response --- $\Delta \varepsilon(t) = J_{\mathrm{GZ}}(t) \Delta\sigma_{l}$ has the explicit solution \cite{Mainardi2011}:
\begin{equation}
        \begin{array}{rl}
    J_{\mathrm{GZ}}(t) &
    =\frac{1}{E}\left[{
    \frac{E}{E_{m}+E} + \frac{E_{m}}{E_{m}+E}
    \left({
    1-\mathbf{E}_{\alpha}\left({-(t/\tau)^{\alpha}}\right)
    }\right)
    }\right]\\
    &\\
   & =\frac{1}{E}\left[{
    1-\frac{E_{m}}{E_{m}+E}
    \mathbf{E}_{\alpha}\left({-(t/\tau)^{\alpha}}\right)
    }\right] ,
\end{array} 
\end{equation}
where
$\mathbf{E}_{\alpha}\left({z}\right) := \mathbf{E}_{\alpha,1}\left({z}\right)=\sum_{n=0}^{\infty}\frac{z^{n}}{\Gamma(\alpha n+1)} $ 
denotes the so-called Mittag-Leffler function which can be evaluated to the limits:  
$\lim_{z\to 0^{+}} \mathbf{E}_{\alpha}\left({z}\right) = 1 - \frac{z}{\Gamma(\alpha+1)}$ and 
$\lim_{z\to +\infty} \mathbf{E}_{\alpha}\left({z}\right) = - \frac{1}{z \Gamma(1-\alpha)}$.
By simplifying the transient term $\mathbf{H}_{\alpha}(t/\tau):=\frac{E_{m}}{E_{m}+E} \mathbf{E}_{\alpha}\left({-(t/\tau)^{\alpha}}\right)$ , the response of each fiber to a sudden increase in the local stress $\Delta \sigma_{l}$ reads: 
\begin{equation}
    E\Delta \varepsilon (t) = (1-\mathbf{H}_{\alpha}(t/\tau))\Delta \sigma_{l}. 
    \label{eq:BD:transientElement}
\end{equation}
%$:=\frac{E_{m}}{E_{m}+E}\Large{\mathbf{E}}_{\alpha,1}\left({- \left({\frac{t}{\tau}}\right)^{\alpha}}\right)$, being $\Large{\mathbf{E}}_{\alpha,1}(x)$ the so-called Mittag-Leffler function. The term $\mathbf{H}_{\alpha}(t/\tau)$ 
Here, the transient term evolves from a positive value:
\begin{equation}
    h:=\mathbf{H}_{\alpha}(0)= \frac{E_{m}}{E_{m}+E}
\end{equation} 
to $\mathbf{H}_{\alpha}(t/\tau \gg 1) \to 0$. 
A sudden increase in the local stress will induce an initial sudden increase in strain of the elastic element $ E \Delta \varepsilon(0^{+})  =  \frac{E}{E+E_{m}}  \Delta\sigma_{l} $ which is lower than in the standard DFBM ($ E \Delta \varepsilon(0^{+})  = \Delta\sigma_{l} $). The rest of the elastic energy is retained by the springpot element and slowly released to the spring element during the creeping phase. This creep response shares similarities with the addition of viscoelasticity to the elastic rebound model proposed in Ref.~\cite{Benioff1951}.
The creeping time of the springpot introduces a third temporal scale to the model, apart from the interaction between fibers and the driving.
This additional temporal scale is responsible for the emergence of temporal correlations in this model (section \ref{sec:BD:trigger}).
We consider that the interactions between fibers are much faster than the relaxation of the springpot. Under quasistatic driving, all temporal scales are much faster than the driving. This implies that the response value for the standard DFBM is reached before the system is driven again, since $\mathbf{H}_{\alpha}(t/\tau \gg 1) \to 0$.\\

\begin{figure}
    \includegraphics[width=1.0\columnwidth]{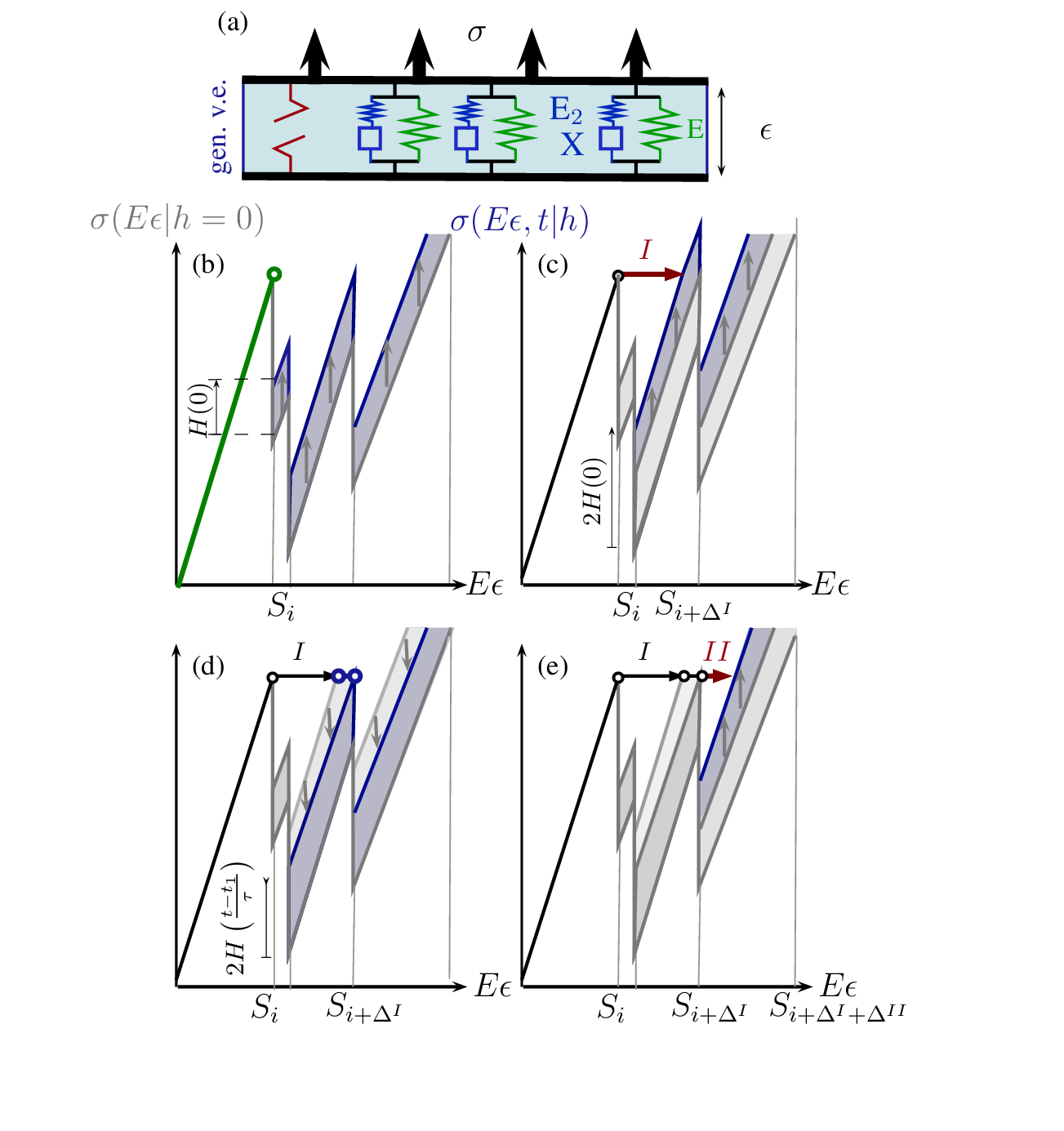}\\ 
\caption{\label{fig:BD:gve} (a) Schematic representation of the generalized rheological democratic fiber bundle model (GVE-DFBM), where the elastic element has been substituted by a generalized Zener solid element. (b--e) Sketch of an avalanche process in the GVE-DFBM with the same fibers considered in Fig.~\ref{fig:BD:std}. (b) The system is hardened a factor $H(0)$ for each failed fiber (blue lines). (c) The initial avalanche, of size $\Delta^{I}$, stops sooner than in the standard DFBM. (d) During the creeping phase and without driving, the hardening is relaxed just enough to activate the next failure. (e) The process is repeated until all the $\Delta$ fibers of Fig.~\ref{fig:BD:std} are broken. All transient terms are relaxed to 0 before resuming the quasistatic driving. While part of the brittle deformation in the standard DFBM model occur as creep in the GVE-DFBM, the number of broken fibers remains the same.}
\end{figure}

As represented, the Zener solid element has three free parameters: $\tau$, $\alpha$ and $h$. 
Within the framework of our model, the generalized relaxation time-scale $\tau$ is arbitrary as we assume a time-scale separation between the quasistatic driving, the relaxation time and the instantaneous avalanche propagation.
The fractional dimension $\alpha$ controls the profile of the relaxation process. By imposing $\alpha=1$ we recover the standard Zener element, where $X:=\eta$ representing a viscous dashpot and the corresponding term $\mathbf{H}_{1}(t/\tau) = h \exp(-t/\tau)$. Lower values of $\alpha$ imply a more complex memory in the relaxation that cannot be simplified in an exponential decay. Instead, the memory is characterized by a power-law decay. Thus, $\alpha$ controls the temporal correlations between avalanches but has a minor role on their size and number.
%The numerical results presented in this manuscript are limited to $\alpha=1$, while all analytical solutions are generalized to any physical value of $\alpha$, as will be shown.
Instead, these are controlled by the hardening parameter $h$. For $\alpha=0$ or $h=0$, equivalent to setting $E_{m}\ll E$, we recover the elastic response. We use this case as a benchmark to the standard DFBM in our numerical simulations. 
For  $E\ll E_{m}$ \cite{Kawada2006}, we recover a generalized Kelvin-Voigt element, with $h=1$. This case imposes continuity in $\varepsilon$ and hence all fibers break individually, since a sudden increase in $\sigma_{l}$ does not generate a sudden stress drop. The implementation of fiber bundle models with Kelvin-Voigt elements is briefly discussed in Ref.~\onlinecite{Kun2006}. Here, we solve analytically the more general viscoelastic DFBM (GVE-DFBM) by using the  mechanical behavior of the individual generalized Zener elements in the constitutive equation of the DFBM (see Fig.~\ref{fig:BD:gve}.a). \\

Considering the stress variations due to fracture in \eqref{eq:BD:transientElement} and equal load sharing ($\sigma_{l}=\sigma/(1-F)$), the constitutive equation of the GVE-DFBM will match \eqref{eq:BD:adiabatic} when all terms $\mathbf{H}_{\alpha}(t/\tau) \to 0$, i.e., on the time-scale of the quasistatic driving. Directly after the breaking of fibers, the constitutive equation depends on the historical sequence of the recent avalanches $\lbrace j \rbrace$ occurring at the frozen value of $\sigma$, and can be expressed as:
\begin{equation}
    E \varepsilon(t) = \sigma(t)\left({ \frac{1}{(1-F(E\varepsilon))} -\sum_{S_{j}<E\varepsilon} \phi_{j}\left({t-t_{j}}\right)}\right) ,
            \label{eq:BD:transient}
\end{equation}
where the terms:
\begin{equation}
  \phi_{j} \left({t-t_{j}}\right) := 
  %\Delta \sigma_{l} \mathbf{H}_{\alpha}\left({\frac{t-t_{j}}{\tau}}\right) = 
  \frac{\delta F_{j} \mathbf{H}_{\alpha}\left({\frac{t-t_{j}}{\tau}}\right)} {(1-F(S_{j+\Delta_{j}}^{-}))(1-F(S_{j}^{-}))}  
\end{equation}
 contain the contribution of each preceding avalanche $j$, with integer size $\Delta_{j}:=M \delta F_{j}$ initiated at strength values $S_{j}^{-}$. 
On the contrary to \eqref{eq:BD:adiabatic}, the constitutive equation has a temporal dependence on the history of the process and cannot be simplified as a function of state. Thus, the avalanche activity rate will exhibit temporal correlations, absent in the standard DFBM. 
%Quasistatic driving is modeled by examining the minimum stress needed to break the next fiber (...) imposing $d\sigma/dt=0$ during the relaxation of the $\phi_{j}$ terms. Hence, the contribution of all the terms  

\subsection{Viscoelasticity as a transient hardening}
\label{sec:BD:trans}

We can prove that the GVE-DFBM is a specific implementation of a more general transient hardening model by taking the thermodynamic limit. 
%We can show that the transient term in Eq.~\eqref{eq:BD:transientElement} acts as an effective transient hardening. 
In the standard DFBM, the instantaneous failure of a system fraction $\delta F$ at $S=E\varepsilon$  introduces a drop in the constitutive Eq.~\eqref{eq:BD:adiabatic} corresponding to 
\begin{equation}
    \delta \sigma_{\mathrm{std}}= - E\varepsilon {\delta F} .
\end{equation} 
In the GVE-DFBM \eqref{eq:BD:transient} the same event will cause a stress change:
\begin{equation}
%    \delta \sigma= - E \varepsilon \delta F (1-\mathbf{H}_{\alpha}(t/\tau))=\delta \sigma_{\mathrm{std}}(1-\mathbf{H}_{\alpha}(t/\tau)){}
%     \delta \sigma=\delta \sigma_{\mathrm{std}}\frac{1-\mathbf{H}_{\alpha}(t/\tau)}{1-\frac{\delta F\mathbf{H}_{\alpha}(t/\tau)}{1-F+\delta F}}
     \delta \sigma=
      \frac{\delta\sigma_{\mathrm{std}} \left({1-\mathbf{H}_{\alpha}(t/\tau)}\right)}{(1-(1-F)\sum \phi)(1-(1-F-\delta F)(\sum \phi+\phi_{s})) } ,
    \label{eq:BD:response}
\end{equation} 
where $\phi_{s}= \frac{\delta F h}{(1-F(S_{j+\Delta}^{-}))(1-F(S_{j}^{-}))}$ is the $\phi$ term due to the latest failure of $\delta F$. We have relabeled the term $F:=F(E\varepsilon ^{-})$. In the thermodynamic limit, $\delta F/(1-F) \ll 1$, and the denominator can be approximated as 1 at the lowest order. Thus, in this limit the mechanical response \eqref{eq:BD:response} reduces to: 
\begin{equation}
\delta \sigma \approx   \delta \sigma_{\mathrm{std}}\left({1-\mathbf{H}_{\alpha}(t/\tau)}\right){}.
\end{equation}
Since the stress drop is reduced with respect to $\delta \sigma_{\mathrm{std}}$, the system regains the stability with a lower deformation than the standard DFBM (see Fig.~\ref{fig:BD:gve}.c--e). Hence, we can interpret the transient term  $\mathbf{H}_{\alpha}(t/\tau)$  in Eq.~\eqref{eq:BD:transientElement} as a transient hardening with respect to the standard DFBM, increasing temporally the effective strength in $\sigma$ of all the surviving elements. \\

\subsection{The origin of aftershocks}\label{sec:BD:as}

The history dependence in the constitutive equation is a mechanism able to generate temporal correlations, that can be expressed in terms of a triggering point process, where each avalanche is either a `background' event activated by the external driving or a `triggered' event when is direct consequence of previous activity.
This is consistent with the event-event triggering or aftershock picture typically invoked for seismic events~\cite{Ogata1988, Gu2013, Davidsen2016} and AE events \cite{Baro2013, Davidsen2017, Baro2017} to explain temporal correlations. Specifically, the rate of events triggered by a given event decays over time, with a typical power-law profile that is consistent with the relaxation of generalized viscoelasticity \cite{Zhang2016}.  
Fig.~\ref{fig:BD:gve}.c--e represents schematically the avalanche process in the GVE-DFBM (blue curve) in comparison to the standard DFBM (gray curve). The breaking of the fibers at $S_{i+\Delta^{I}}$  in Fig.~\ref{fig:BD:gve}.d, retarded by effect of hardening, occurs at the same stress $\sigma$ that triggered the primary avalanche (at $S_{i}$ in Fig.~\ref{fig:BD:gve}.b), since we consider that the relaxation time needed to activate the secondary avalanche is much faster than the quasistatic driving. Thus, the driving is not directly responsible of the secondary avalanche in  $S_{i+\Delta^{I}}$. Instead, it is the failing of the elements broken in the avalanche at $S_{i}$ (this one due to the driving) which triggers the failure. Thus, we can classify the events into background (event $I$ starting at $S_{i}$ in this example) and triggered (event $II$ starting at $S_{i+\Delta^{I}}$ in this example).  \\

While the temporal evolution has changed on the time scale of the relaxation, the $\sigma (\varepsilon)$ diagram is invariant to transient hardening. As a consequence, considering the same driving conditions, a given avalanche in the standard DFBM is split into a cluster of causally correlated avalanches in the GVE-DFBM. When all the fibers forming the avalanche in the standard DFBM -- now, the cluster -- have been broken, all terms $\phi_{j}(t)$ are relaxed to zero. This imposes temporal independence between clusters, since \eqref{eq:BD:transient} is equivalent to \eqref{eq:BD:adiabatic} in that case. \\

Since the interactions are mean field, all correlations between avalanches are determined by a scalar relation between the activation strengths. Independently of the value of $h$, an individual fiber with strength $S_{j}$ will break as a consequence of a previously broken fiber $S_{k}$ if $ S_{j}(1-F(S_{k})-1/M) < \sigma < S_{j}(1-F(S_{k}))$. Depending on the value of $h$, the breaking of $S_{j}$ will occur either within the same avalanche or within a latter triggered avalanche (ie. aftershock) within the same cluster, only when $h>0$. 
Hence, mean field models are unable to generate the superposition of complex triggering trees identified in natural phenomena \cite{Baiesi2004,Zaliapin2013,Gu2013,Hainzl2016} and modeled in spatio-temporal stochastic point processes \cite{Ogata1988,Turcotte2007,Davidsen2016}. Instead, all aftershocks triggered due to the breaking of a given fiber $S_{k}$ are correlative in time and occur in the same cluster. However, even the MF approximation is able to render the power-law temporal statistics (see section \ref{sec:BD:trigger}), supporting the link between triggering process and the phenomenological observations of aftershocks in AE experiments and seismicity. From the analytical results derived in the following sections, we can argue that the details regarding the structure of the triggering trees shall not have a significant impact on the shape of the avalanche size distribution in the thermodynamic limit.
\\

\section{The mean field universality class for fracture with transient hardening.}
\label{sec:BD:MF}

As mentioned in section~\ref{sec:BD:as}, the stress value for an avalanche to occur in the standard DFBM coincides with the stress value of a cluster in the GVE-DFBM. The constitutive curves of both models are indistinguishable in the thermodynamic limit, and so is the coarse grained effective modulus $G=d\varepsilon/d\sigma$ as well as the moment released per time unit $d\Delta/dt$. However, the number of avalanches and their statistical properties have been strongly altered. Avalanches tend to be smaller (Eq.~\ref{eq:BD:response}) due to the effect of hardening  and yet, as we will prove now, all avalanche statistics fall inside the same universality class, regardless of the value of $h$. \\

\begin{figure}
\includegraphics[width=\columnwidth]{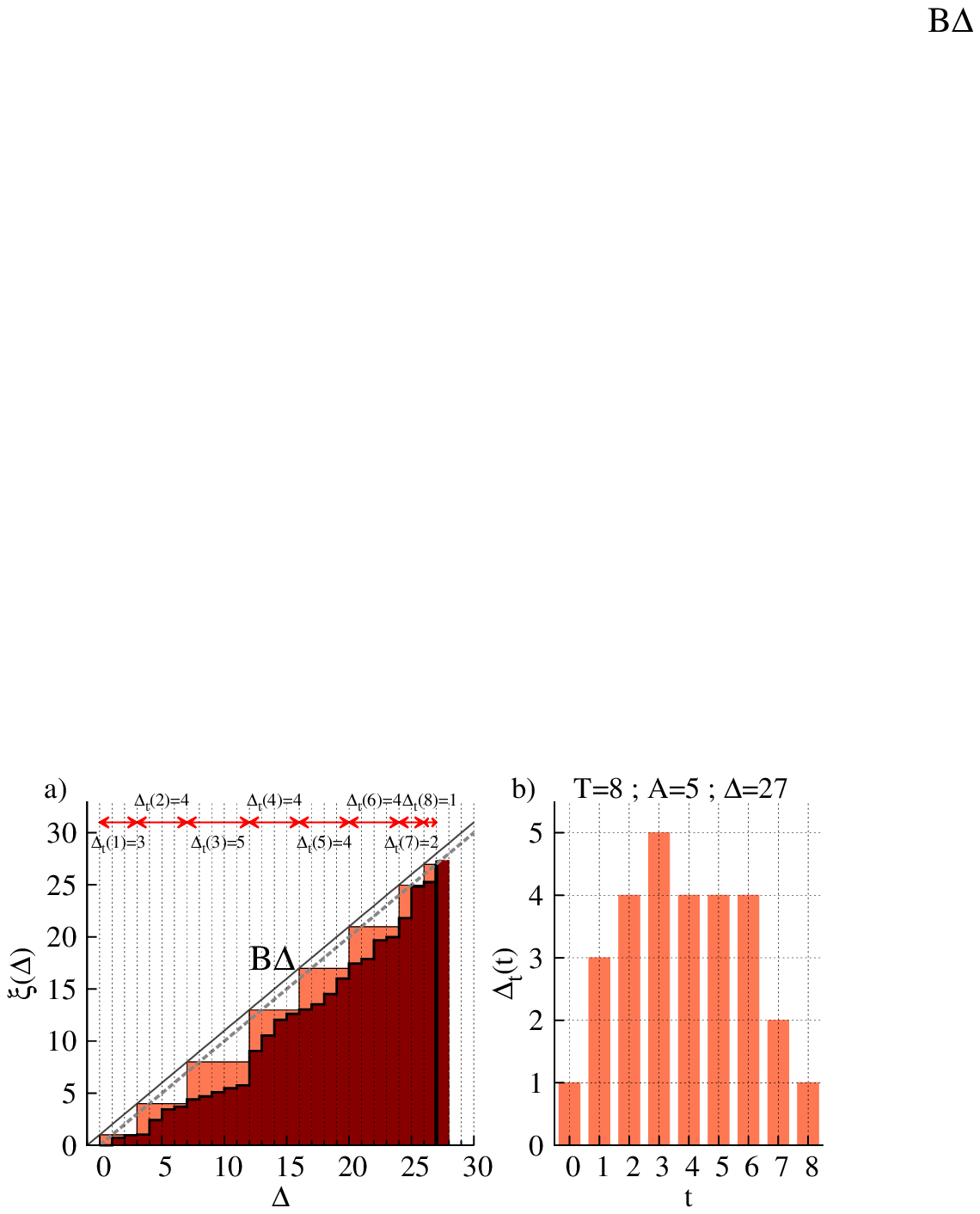}\\
\includegraphics[width=\columnwidth]{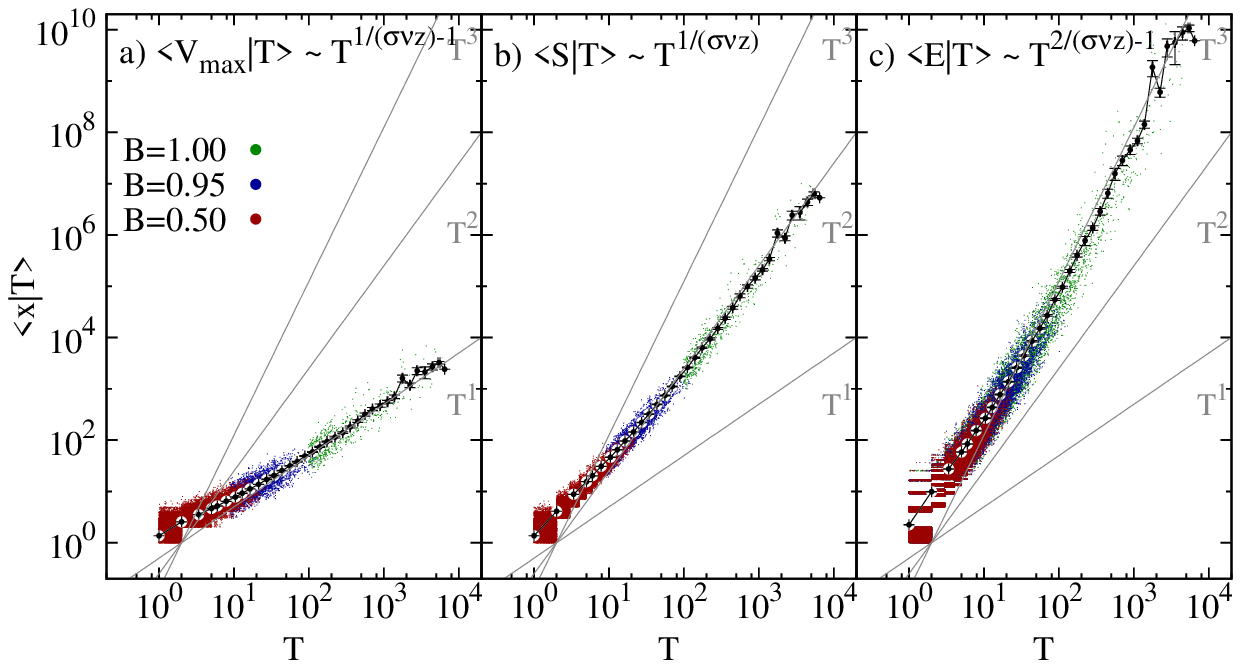}\\
\caption{\label{fig:BD:hitTimes} (a) Example of an avalanche defined as the hitting times of a stationary random counting process $\xi(\Delta)$ to the boundary $B \Delta$. Maroon area shows the landscape $\xi(\Delta)$ drawn from random Poisson increments at each elementary step $\Delta$. The avalanche stops at the first $\Delta$ value with $\xi(\Delta)> B \Delta$ represented as a dashed gray line. The salmon area represents the elements failing at the time unit, with values $\xi(\Delta)< B(\Delta(t-1) + 1)$ (black solid line). (b) The temporal profile of the same avalanche, defining the amplitude $A$, duration $T$ and size $\Delta$. (c--e) Scatter-plots of avalanches \cite{footnote1}
in the space (c) $A,T$, (d) $S,T$, (e) $E,T$ found for different values of $B$ and the conditional averages $\langle x | T \rangle$ for $B=1$ (black error lines). }
\end{figure}

 Under quasistatic driving, an avalanche  starting at the failure of fiber $i$ with strength $S_i$ will stop at the first fiber $i+\Delta_{i}$ with strength $S_i+\delta s$ such that: $\sigma(S_{i+\Delta_{i}})>\sigma(S_{i})$. Thus, the size of the avalanche $\Delta_{i}$ is defined as the number of broken fibers from a process of record dynamics and related to the difference in the fraction of broken fibers ($\delta F := F_{i+\Delta}-F_{i}$)  as  $\delta F \equiv \Delta/M$. Since the propagation of the avalanche is much faster than the viscoelastic relaxation, the contribution to the transient term of all the fibers broken within the same avalanche is $\phi_{i}=\frac{ \delta F \mathbf{H}_{\alpha}(0)}{(1-F_{i})(1-F_{i}+\delta F)}$.
 From the constitutive equation \eqref{eq:BD:transient}, the avalanche stops when:
\begin{equation}
\frac{\delta s}{\delta F} > \frac{S_{i}\left({1- \mathbf{H}_{\alpha}(0)}\right)}{(1-F_{i}-\delta F)}\frac{1}{(1-(1-F_{i})\sum_{j} \phi_{j}(t-t_{j}))} .
\label{eq:BD:aval}
\end{equation}
The right-hand side of the equation is constituted by two terms. The term: $\frac{1}{(1-(1-F_{i})\sum_{j} \phi_{j}(t-t_{j}))}$ contains the effect of previous avalanches on the size of the current one and is static during the propagation of the avalanche thanks to the separation between temporal scales. Furthermore, in the limit of small avalanches compared to the system size, this term can be approximated by 1. The term $\frac{S_{i}\left({1- \mathbf{H}_{\alpha}(0)}\right)}{(1-F_{i}-\delta F)}$ includes a depenence on the current size (in $\delta F$) of the avalanche. This dependence is related to the indetermination of the state of the system when the avalanche is large. In the limit of small avalanches: $1-F_{i}-\delta F \sim 1-F_{i}$ wich defines the state of the system. Let's take, for now, the small-avalanche approximation as valid. The right-hand side of \eqref{eq:BD:aval} is simplified as a function of $S_{i}$:
\begin{equation}
 b(S_{i}|h) := \frac{S_{i}}{1-F_{i}} \left({1- h}\right) .
\end{equation}
The left-hand side can be redefined in a dimensionless form. By definition: ${\delta F}:= \Delta / M $. Since the values of $S_{i}$ are i.i.d., the increment $\delta s$ between $\Delta$ consecutive strengths is a Poisson process of $\Delta$ trials at rate $M p(s)$, where $p(s)$ is again the probability density function of $S_{i}$ values. The dimensionless form of \eqref{eq:BD:aval} for this general representation of the transient hardening model reads:
\begin{equation}
 \frac{\xi(\Delta_{i})}{\Delta_{i}} > B(S_{i}|h) ,
 \label{eq:BD:aval2}
\end{equation}
as represented in Fig.~\ref{fig:BD:hitTimes}.a, where $B(S_{i}|h)  := p(S_{i}) b(S_{i}|h)$ and $\xi := M p(s)  \delta s$ is a Poisson process of rate $1$. 
Given a single realization of strengths at fixed $h$, once an avalanche has started at $S_{i}$, the value $B(S_i)$ acts as a constant threshold and the distribution of avalanches in the GVE-DFBM is equivalent to the distribution of first hitting times of a random Poisson counting process $\xi(\Delta)$ to the moving boundary $B(S_{i})\Delta$. \\

For $B>1$  there is a macroscopic probability that an avalanche grows to an infinite size. We associate this supercritical regime to a brittle failure event. For $B<1$ the probability of an infinite avalanche is $0$, and the distribution of finite avalanche sizes can be approximated as a generalized homogeneous function:
\begin{align}
 D(\Delta;B)d\Delta & 
 = \Delta^{-\kappa_{\Delta}} \; \mathcal{D}(\Delta |1-B|^{\beta_{B}}) \;d\Delta\\
 & = |1-B|^{\kappa_{\Delta} \beta_{B}} \;  \widetilde{\mathcal{D}}(\Delta |1-B|^{\beta_{B}}) \;d\Delta ,
 \label{eq:BD:distro}
\end{align}
where $\kappa_{\Delta}$ and $\beta_{B}$ are universal exponents, $\widetilde{\mathcal{D}}(\Delta |1-B|^{\beta_{B}}) := (\Delta |1-B|^{-\beta_{B}})^{\kappa_{\Delta}} \mathcal{D}(\Delta |1-B|^{\beta_{B}})$ and  $\mathcal{D}(\Delta |1-B|^{\beta_{B}})$  are  scaling functions.
This scaling term diverges exactly at $B_{c}=1$, which defines a critical point with scale-free avalanches. It is important to remember that $B$ is constant only in the regime of small avalanches. In both the standard DFBM and GVE-DFBM, this limit can be achieved asymptotically close to a critical point --- where avalanches are scale-free --- by increasing the size of the system. In the thermodynamic limit ($M \to \infty$), the yielding process up to the critical point is well defined by \eqref{eq:BD:aval2}. But, according to \eqref{eq:BD:aval}, at criticality and above, when $B(\delta F \to 0) \geq 1$, the avalanche might grow to sizes such that $B(\delta F \ ) > B(\delta F \to 0)$ and, thus, the system is, by definition, supercritical. Although $B=1$ is critical, the critical point in the DFBM is not well defined because of the coupling between the the avalanche size ($\Delta$) and the state of the system ($B$). From now on, the reader shall keep in mind that the expression \eqref{eq:BD:aval2} and the following derivations are valid for $B<1$ in the DFBM.\\

 \begin{figure}
    \includegraphics[width=1.0\columnwidth]{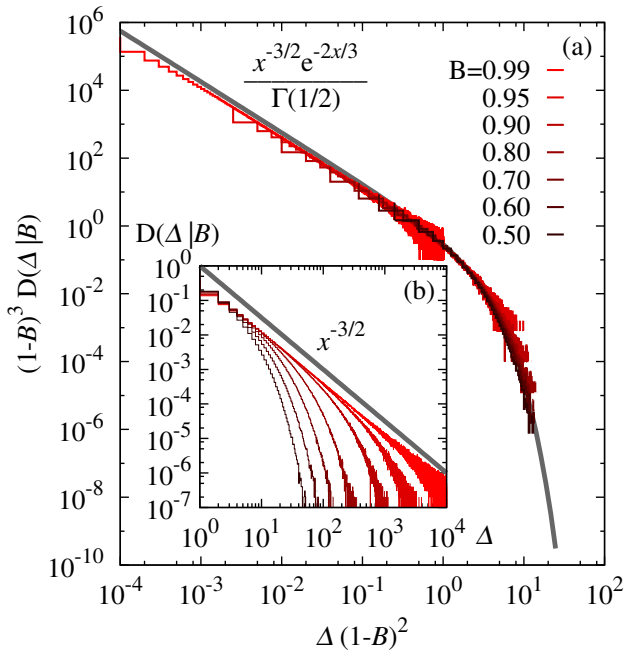}\\ 
\caption{\label{fig:BD:pHit} Distribution of $N=10^{7}$ hitting times for the Poisson process $\xi_{\Delta}$ to the boundary $B \Delta$ tapped at $\Delta=10^{4}$ for different values of $B$. (a) Scaling according to \eqref{eq:BD:distro} compared to the ansatz (grey thick line) in \eqref{eq:BD:ansatz}. (b) Distribution before scaling, compared to the power-law expected by $B=1$ (gray thick line).}
\end{figure}
 
Fig.~\ref{fig:BD:pHit}.b shows the numerical distribution of return times ($\Delta$) for different values of $B$. The probability distribution functions collapse onto a single universal function given the scaling relations with $B$ stated in \eqref{eq:BD:distro}, as shown in Fig.~\ref{fig:BD:pHit}.a. The fitted critical exponents are $\kappa_{\Delta}=3/2$ and $\beta_{B}=2$. The exponent $\kappa_{\Delta}=3/2$ is ubiquitous in the distribution of avalanche sizes in mean field models. The exponent $\beta$ is usually defined as a function of the driving mechanism and the relation with $B$ has to be derived, as shown in the next section for the case of the DFBM. \\

Since \eqref{eq:BD:aval2} is dimensionless, all the information, including the distribution of avalanches, is fully determined by the scalar term $f_{B}:=1-B$ measuring the distance a critical point. In the case of the transient hardening model, the value of $B$ is defined in the thermodynamic limit given a hardening $h$, a strain value $s/E$ and the sampling distribution of $S_{i}$. Since the functional form of $\mathcal{D}$ is invariant to the explicit dependence of $B$ with the state of the system, any model that can be represented as \eqref{eq:BD:aval2} fulfills the scaling relation Eq.~\eqref{eq:BD:distro}. 
Considering the distribution represented in Fig.~\ref{fig:BD:pHit}.a the specific functional form of $\mathcal{D}$ can be approximated to the ansatz:
\begin{equation}
\mathcal{D}(x)=\exp(-3x/2)/\Gamma(-0.5) .
\label{eq:BD:ansatz}
\end{equation}
However, one can show that this approximation is inadequate to measure some quantities such as $\langle \Delta \rangle$, specially for $B\ll 1$, due to the discrete nature of $\Delta$. Instead,  we use the following ansatz for the dependence of the statistical moments  on $f_{B}$: 
\begin{equation}
\langle \Delta^{n} | f_{B}\rangle = f_{B}^{(\kappa_{\Delta}-1-n)\beta_{B}}.
\label{eq:BD:ansatz2}
\end{equation}
Therefore, \eqref{eq:BD:ansatz2} will replace \eqref{eq:BD:ansatz} when possible to compare analytical and numerical results. 
As a consequence of this power-law relation, even if the explicit dependence of a model on $B$ is unknown, \eqref{eq:BD:distro} can be rewritten in terms of the first statistical moment $\langle \Delta | f_{B} \rangle = f_{B}^{(\kappa_{\Delta}-2)\beta_{B}} $ as : 
\begin{equation}
D(\Delta;\langle \Delta \rangle)d\Delta = \langle\Delta \rangle^{\dfrac{\kappa_{\Delta}}{\kappa_{\Delta}-2}} \widetilde{\mathcal{D}}(\langle\Delta \rangle^{\dfrac{1}{\kappa_{\Delta}-2}} \Delta) d\Delta    
\end{equation}
for a fixed $f_B$. This expression depends only on the specific exponent $\kappa_{\Delta}$ with MF values $\frac{\kappa_{\Delta}}{\kappa_{\Delta}-2} = -3$ and $\frac{1}{\kappa_{\Delta}-2}=-2$.  Notice that this expression is more general than \eqref{eq:BD:distro} and may be also fulfilled by other models incompatible with \eqref{eq:BD:aval2}.\\

In order to fully characterize the universality class (UC), we define a time unit within the temporal scales of avalanche propagation and associate a temporal profile to the avalanche propagation by designating the causality tree-like structure between the failing fibers.  In terms of a DFBM, the breaking of an original fiber at time unit 0 can cause the breaking of a number fibers during time unit 1. Such fibers will cause the breaking of other fibers at time unit 2, etc. The temporal profile at time $t$ is determined by the number $\Delta_{t}^{i}$ of fibers with associated values $\xi({\Delta(t-1) + \Delta_{t}^{i}}) < B(\Delta(t-1) + 1))$, i.e. all the values of $\xi$ that can be activated by the state of the system at time $t-1$. In Fig.~\ref{fig:BD:hitTimes}.a, the salmon areas illustrate the set of fibers breaking together in a time unit (until $\xi({\Delta(t-1) + \Delta_{t}^{i}})$ hits the value $B(\Delta(t-1) + 1))$ given by the black solid line). The intensity of the temporal profile is represented with arrows in Fig.~\ref{fig:BD:hitTimes}.a, and histograms in Fig.~\ref{fig:BD:hitTimes}.b. Apart from the avalanche size $\Delta^{i}=\sum_{t=0}^{T} \Delta_{t}^{i}(t) $, the temporal profile $\Delta_{t}^{i}(t)$ allows to define additional variables: a duration $T^{i}$, as the number of time units; an amplitude $A^{i}$ corresponding to $\max(\Delta_{t}^{i}(t))$; and also an energy $E^{i}=\sum_{t=0}^{T} \left({\Delta_{t}^{i}(t)}\right)^{2}$, usually related to the seismic release and acoustic emission measurements in the literature \cite{Dahmen2017}.\\

Assuming that the process defining the hitting times is scale-invariant over a broad range of scales, the average avalanche profile must scale with the duration such that: 
\begin{equation}
    \langle \Delta_{t} (t) | T \rangle = T^{\frac{1}{\sigma \nu z} -1} \Phi(t/T) .
    \label{eq:BD:scaleInvariance}
\end{equation}
The average relation between the four magnitudes can be summarized as:
\begin{equation}
    \langle A | T \rangle \sim T^{\frac{1}{\sigma \nu z} -1}  \quad 
    \langle \Delta | T \rangle \sim T^{\frac{1}{\sigma \nu z}}  \quad
    \langle E | T \rangle \sim T^{\frac{2}{\sigma \nu z} -1} . 
\end{equation}
The numerical results of the conditional averages are shown in Fig.~\ref{fig:BD:hitTimes}.c--e. Although the density distributions depend on $B$, the average relationships between magnitudes is conserved and agrees with \eqref{eq:BD:scaleInvariance} given a value $\sigma \nu z = 1/2 $, coinciding with the mean value for stick-slip models \cite{Dahmen2011}.

\section{Simulation results}
\label{sec:BD:num}

\subsection{Interpretation of `B' in terms of the driving in the standard and GVE-DFBM}
\label{sec:BD:B}

In each specific model of critical failure, the particular exponent $\beta$ associated to the distance to failure is determined by the explicit relation between $B$ and the mechanism of external driving such as a constant stress ($\sigma$) or strain ($\varepsilon$) rate driving. In such cases, we can formulate the time to failure in terms of distance to the macroscopic failure point in strain: $E\varepsilon_{f}= \frac{1-F(E\varepsilon_{f})}{p(E\varepsilon_{f})}$, or stress: $\sigma_{f} = \frac{(1-F(E\varepsilon_{f}))^{2}}{p(E\varepsilon_{f})}$.
In the case of the GVE-DFBM, the general relation of $B$ with strain can be obtained to a good approximation in the thermodynamic limit by expanding $B$ around the failure point:
\begin{equation}
   B = (1-h)\left[{1 - f_{\varepsilon}\left({
   2+{\varepsilon_{f}}\left.{\frac{d}{d \varepsilon} \log(p(E\varepsilon))}\right|_{\varepsilon_{f}}
   }\right) + O(f_{\varepsilon}^{2}) }\right],
   \label{eq:BD:esigmaB}
\end{equation}
where $\mathit{p}(E\varepsilon)$ is the strenght ($S_{i}$) distribution evaluated at $E\varepsilon$. This relation is linear in a first order approximation. The strength sampling distribution only affects the constant term $2+{\varepsilon_{c}}\left.{\frac{d}{d \varepsilon} \log(p(E\varepsilon))}\right|_{\varepsilon_{c}}=:2A$. As specific cases, if $S_i$ is uniformly distributed, $A=1$, while for a Weibull distribution, $A=1+m/2$.
  Considering the first order approximation \eqref{eq:BD:esigmaB}, $f_{B}\approx 2(1-h)f_{\varepsilon}A + h$ and, at the yield point (when $f^{f}_{\varepsilon}=0$), $f^{f}_{B}\approx  h$. Critical failure only occurs for $h=0$, corresponding to the standard DFBM. The model with $h=0$ is critical in terms of \eqref{eq:BD:critFail} with the exponent $\beta_{\varepsilon}=2$. As a particular result, we notice that the characteristic scale at failure ($f_{\varepsilon}=f_{\sigma}=0$) scales with $h$ as: $\langle \Delta | h \rangle \sim h^{-1}$. Instead, the MF solution of stick-slip models reports a scaling $\langle \Delta | \epsilon \rangle \sim |\epsilon|^{-2}$  \cite{Mehta2006}, with $h$ being equivalent to $-\epsilon$. The discrepancy in this exponent is discussed in the next section.\\

We can find the relation with stress ($\sigma$) by expanding the constitutive equation around $\varepsilon$ close to the critical point. Under quasistatic driving, $\sigma(E\varepsilon)$ is equivalent to Eq.~\ref{eq:BD:adiabatic} and around $\varepsilon_{f}$ reads:
\begin{equation}
   f_{\sigma}=f_{\varepsilon}^{2}\left({
   1+\frac{\varepsilon_{f}}{2}\frac{d}{d \varepsilon} \log(p(E\varepsilon_{f}))
   }\right) + O(f_{\varepsilon}^{3}) .
   \label{eq:BD:fsigmaB} 
\end{equation}
Thus, $f_{\sigma}\approx A f_{\varepsilon}^{2}$ and $f_{B} \approx (1-h)\left({A f_{\sigma}}\right)^{1/2} + h$. For the standard DFBM, we find critical failure (Eq.~\eqref{eq:BD:critFail}) with $\beta_{\sigma}=1$, as expected from the mean field solution of the standard DFBM \cite{Pradhan2002}.
As an example, for a Weibull distribution and standard ($h=0$) conditions: $f_{B}(f_{\varepsilon})=1-(1-f_{\varepsilon})$ and $f_{B}(f)=1+ W\left({-\frac{(1-f_{\sigma})^{m}}{e}}\right)$ where $W(x)$ is the Lambert function, inverse of $x=W\exp(W)$. We can expand $f_{\sigma}$ in terms of $f_{B}$ by inverting this expression: 
\begin{equation}
f_{\sigma}=1+(f_{B}-1)\exp(f_{B})= \frac{f_{B}^{2}}{2m}+\frac{f_{B}^{3}}{3m}+O(f_{B}^{4})  ,
\label{eq:BD:fsigma}
\end{equation}
 thus satisfying the approximate relation $f_{\sigma} \sim f_{B}^{2}$, as expected based on the approximations in \eqref{eq:BD:fsigmaB} and \eqref{eq:BD:esigmaB}. \\

\subsection{Distribution of avalanche sizes}
\label{sec:BD:distro}

\begin{figure}
\includegraphics[height=\columnwidth, angle=270]{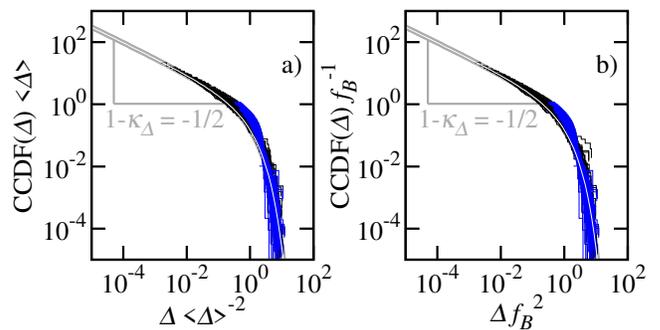}\\
\caption{\label{fig:BD:dD} Complementary cumulative distribution function (CCDF) of avalanche sizes scaled by (a) their mean value ($\langle \Delta \rangle$) and (b) distance to criticality ($f_{B}$), obtained by the numerical simulations of the GVE-DFBM with $m=1$ (std. Zener elements) and $M=10^{7}$ evaluated in intervals of $\sigma$. We compare the results for $h=0.4$  (in blue) with the standard DFBM ($h=0$ in black) and the universal distribution for the analogous hitting time problem according to \eqref{eq:BD:ansatz} (gray thick lines). }
\end{figure}

Given the distribution of $\Delta$ \eqref{eq:BD:distro} and considering the relation between $\sigma$ and $B$ derived from \eqref{eq:BD:fsigmaB}  and \eqref{eq:BD:esigmaB}, we can forecast the expected distribution of sizes, durations, amplitudes and energies for the GVE-DFBM model as function of the distance to the critical point under stress driving, $f_{\sigma}$. As an specific case, the distribution of avalanches sizes, matching the results represented in Fig.~\ref{fig:BD:pHit},   for the standard DFBM will depend explicitly on the distance to the failure stress as:
\begin{equation}
 D(\Delta;f_{\sigma})d\Delta = \Delta^{-3/2} \; \mathcal{D}(\Delta f_{\sigma} ) \;d\Delta{}
 \label{eq:BD:distroF}
\end{equation}
and, thus, differ from the mean field solution for stick-slip models, where the characteristic function scales with $\Delta f_{\sigma}^{2}$. We have shown that this specific exponent --- usually referred to as $1/\sigma$ in the literature --- depends on the relation $B(\sigma)$. Unlike fracture models, stick-slip models restitute or `stick' failed elements, giving rise to a characteristic stationary flow regime under strain driving. If one were able to express the MF stick-slip model in terms of \eqref{eq:BD:aval2}, the relation $B(\sigma)$ would differ from \eqref{eq:BD:fsigma} because of that.\\

 Fig.~\ref{fig:BD:dD} shows the scaling in both $f_{B}$ and $\langle\Delta \rangle$ of the numerical cumulative distribution $\mathrm{CCDF}(\Delta)$ for $h=0$ and $h=0.4$ in a DFBM with a Weibull sampled strengths $s_{i}$ with $m=1$ and $\alpha=1$ (see Appendix for simulation details). The results fit well the normalized ansatz \eqref{eq:BD:ansatz} for the UC \eqref{eq:BD:aval2} as a solution to \eqref{eq:BD:distro}. As expected, Fig.~\ref{fig:BD:dD}.b showing the scaling factor with $f_{B}$ deduced from \eqref{eq:BD:fsigma} is almost indistinguishable to Fig.~\ref{fig:BD:dD}.a showing the scaling with $\langle \Delta \rangle$. This result confirms that the predictions derived from the UC in section \ref{sec:BD:B} are valid in the case of the GVE-DFBM.\\

\subsection{Subcritical failure and foreshocks}
\label{sec:BD:foreshocks}

\begin{figure}
\includegraphics[height=0.95\columnwidth,angle=270]{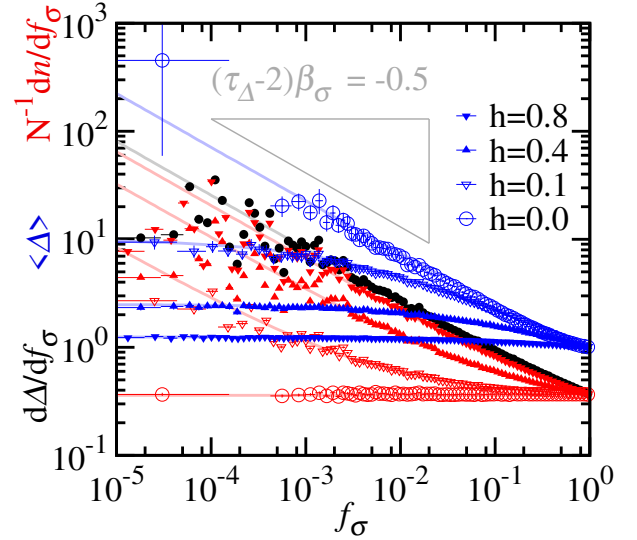}
\caption{\label{fig:BD:scale} Normalized activity rate (red) as the number of events for unit of time ($N^{-1} dn/df_{\sigma}$), average avalanche size ($\langle \Delta \rangle$ in blue) and number of fibers failed for unit of time $d\Delta/dt$ for the GVE-DFBM with values of $h=0.0$, $0.1$, $0.4$ and $0.8$. Light lines serve as a guide to the eye with the analytical solution found for the thermodynamic limit. The temporal scale is expressed in terms to the distance to failure in stress, in order to empathize the agreement with the power-law divergences at the failure point.}
\end{figure}

Thanks to the explicit evolution of $f_{B}$ in \eqref{eq:BD:distro}, we can provide an explanation for the observed lack of divergence in the mean avalanche magnitudes --- either amplitude, size or energy --- in processes exhibiting accelerated seismic release (ASR) proportional to the yielding in the constitutive equation \eqref{eq:BD:transient}. While the number of broken elements over time ($d\Delta/dt (f) = \sum \langle \Delta \rangle (f) dn/dt(f) $) is independent of the rheology in the quasistatic and thermodynamic limit, the evolution of $dn/dt(f)$ and $\langle \Delta \rangle (f)$ will depend on the value of $h$. 
Fig.~\ref{fig:BD:scale} shows the evolution to failure of the numerical results of the GVE-DFBM for different values of $h$. Each data set corresponds to a single simulation for a bundle with $M=10^{7}$ standard ($\alpha = 1$) viscoelastic elements and a fixed $h$. No major differences are expected for other values of $\alpha$ since the results are equivalent in the thermodynamic limit. The strengths are sampled from a Weibull distribution with $m=1$. The results for the standard DFBM with elastic (instead of viscoelastic) elements are represented as $h=0$ (circles). Thick light lines represent the analytical solutions found by the approximation to the thermodynamic limit, exhibiting a good agreement with the simulation results. The expression of $\langle \Delta \rangle (f|h)$ is obtained from the ansatz to $\langle \Delta | B \rangle$  introduced in \eqref{eq:BD:ansatz2} and considering the analytical relation $f_{B}(f_{\sigma})$ expected for the strength distribution.\\

Instead of critical failure, the ultimate failure point in the viscoelastic model is reached at $B<1$, i.e. failure is subcritical. The invariance of $d\Delta/dt$ imposed by the constitutive equation implies a divergence in the activity rate with an exponent that shall match the divergence in $d\Delta/dt$ and the equivalent critical failure for $h=0$, since the power decomposition of $\langle \Delta \rangle$ for $h>0$ has a zeroth order (constant) term. Notice that, strictly speaking, due to the divergence in the activity rate, the associated temporal scales introduced by the viscoelasticity can overlap with the driving, even in the thermodynamic limit. This will distort the approximations taken to obtain \eqref{eq:BD:aval2} and return an avalanche set that may differ from the UC in real systems where the stress evolution is not strictly quasistatic. We don't discuss here the properties of the post-peak activity that may appear as consequence of the splitting of the brittle event in aftershock sequences. We expect this collection of events to fall outside the UC, since the terms $\phi_{j}$ in  \eqref{eq:BD:aval} cannot be neglected any longer.\\

\subsection{Presence of power-law temporal correlations}
\label{sec:BD:trigger}

\begin{figure}
\includegraphics[width=1.0\columnwidth]{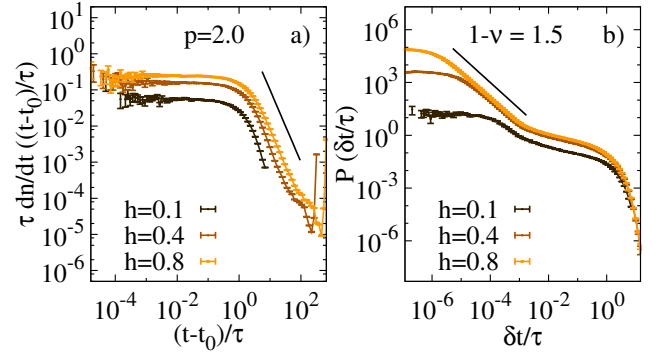}
\caption{\label{fig:BD:as} Temporal correlations in a Zener DFBM (ie. GVE-DFBM with $\alpha=1$) for different values of $h$. (a) Activity rate as function of the time ($t_{0}$) since the beginning of the cluster and (b) distribution of waiting times between consecutive events within a cluster. Time is given in units of $\tau$. Straight lines are added as a guide to the eye illustrating the power-law regimes.}
\end{figure}

Finally, we can verify that, even for standard viscoelasticity ($\alpha=1$), the activity rates observed within clusters are compatible with the Omori relation \cite{Utsu1995} observed in aftershock sequences, reinforcing the link between the presence of aftershocks and the lack of criticality in the presence of transient hardening. 
Fig.~\ref{fig:BD:as} shows (b) the distribution of waiting times ($\delta t$) between events within the same cluster and (a) the apparent decay of the activity rate ($dn/dt$) at time $t$ since the beginning of the cluster at $t_{0}$. The simulations correspond to the simple case $\alpha=1$ and different values of $h$. For high values of $h$, the activity exhibits a power-law regime: $dn/dt(\Delta t) \sim \Delta t ^{-p}$, with exponent $p\approx 2.0$ for $\Delta t \gtrsim \tau$, resembling the modified Omori relation. The distribution of waiting times ($\delta$) exhibits also a power-law regime: $P(\delta)\sim \delta^{-(1-\nu)}$ superimposed onto an exponential distribution. This exponent $1-\nu=1.5$ found in the distribution of waiting times agrees with the relation  $1-\nu=2-1/p$ \cite{Utsu1995} expected if $p$ is the exponent of the triggering kernel (see Ref.~\cite{Corral2003}). All the non-trivial temporal profiles tend to vanish for $h \to 0$, as expected in the limit without temporal correlations ($h=0$) corresponding to the standard DFBM. We expect both exponents $p$ and $\nu$ to be sensitive to the fractional exponent $\alpha$ in generalized implementations \cite{Zhang2016}. The overall distribution of the waiting times and its relation with the triggering rates are particular results of the parametrization, as will be analyzed in future works.

\section{Discussion and conclusions}
\label{sec:BD:conclusions}

The present manuscript provides a plausible relation between the macroscopic observation of temporal correlations and lack of critical failure with a microscopical fundamental principle: the presence of a transient hardening mechanism. The generalized viscoelastic democratic fiber bundle model (GVE-DFBM) serves as an example derived from physical principles of a more general category of variations of the DFBM with some mechanism generating transient hardening. In this explicit model, the amount of hardening is quantified and linked to the observable rheological properties of the material.\\

As a consequence of the transient hardening, the failure point is not critical as one would expect in common conceptual micromechanical models, including the standard DFBM. Instead, the statistical properties of fracture avalanches at the failure point correspond to a subcritical regime with finite correlation lengths and characteristic scales. The activity rate increases up to a divergence compatible with $d\Delta/dt$, which is imposed by the common constitutive equation with the standard DFBM and, thus, is invariant to transient effects under quasistatic driving. One of the most remarkable results is the existence of universal behavior invariant to the parametrization of the model, thus including the standard DFBM. Despite the apparent statistical differences, all the avalanches in any model of fracture compatible with \eqref{eq:BD:aval2} fall within the same universality class (UC), and are only characterized by the distance to the critical point.\\

Notice that this universality class, determined by the reduction of the GVE-DFBM to the hitting times of a counting process \eqref{eq:BD:aval2}, is not exclusive to the implementation of viscoelasticity, nor transient hardening, nor even fiber bundle models. The universality class will be common to any other mean field (MF) lattice models that can be expressed as \eqref{eq:BD:aval2} with any alternative temporal evolution of $B$ or different explicit relation $B=g(S_{i})$. As a particular case, one might expect that the results discussed in the current work can be extrapolated to the incorporation of generalized viscoelasticity to variations of the DFBM such as continuous damage models. 
Furthermore, the statistical properties arising from the representation of the avalanche as a hitting time problem \eqref{eq:BD:aval2} are consistent with other MF UC such as the branching process approach \cite{Zapperi1995} with the same $\tau=1.5$ and also invariant to dissipation \cite{Lauritsen1996}.  A deeper relationship, or even the possible equivalence between the two MF models is yet to be discussed.\\

Interactions in natural fracture processes are anisotropic and have a finite range generating spatially correlated heterogeneities that can lead to nucleation phenomena, macroscopic defects or localization bands. In addition, it is difficult to assess how close a system is to failure at the onset of data recording. 
However, some of the fundamental predictions of this mean field model can be validated by experimental observations. 
The stationarity in the statistical properties of AE events recorded during certain experiments \cite{Baro2013,Castillo2013,Nataf2014} is compatible with the lack of criticality represented in Fig.~\ref{fig:BD:scale} if the natural internal structure of the material is already close to a critical state at the beginning of the experiment. This condition is supported by the wide range of the scale invariance \cite{Baro2013} observed in the stationary energy distribution. Strictly speaking, the amount of AE energy released, and the ASR, will decrease by effect of viscoelasticity due to the energy dissipated by creep. However, in the GVE-DFBM the proportion of dissipated energy is stationary and won't affect the temporal statistics of ASR, which is also a reasonable assumption in more realistic models. 
In contrast, this model cannot provide an explanation to the increase of activity close to failure observed in absence of aftershocks \cite{Davidsen2017}. In these experiments, a link is discussed between temporal correlations and local stress fluctuations emerging due to the presence of large heterogeneities. Such experiences might highlight the role of other processes neglected in this study, such as the weakening of the material due to stress corrosion or the interaction between defects \cite{Kachanov1994, Yamashita1987}.\\

Another phenomenon related to failure that in principle could explain the increase of the energy released is the decrease in the power-law exponent of magnitudes or energies ($\epsilon$ in \eqref{eq:BD:critFail}) sometimes observed close to failure in AE experiments \cite{Scholz1968,Main1989,Goebel2013}. 
Neither conceptual nor numerical micromechanical models of critical failure can reproduce this effect \cite{Amitrano2012}. It has long been suggested that the decrease of the exponent is linked to variations of the stress level \cite{Scholz1968}, a concept that can be related to the distance to failure \cite{Amitrano2012}.
Although not explicitly investigated, the same rheological picture presented in this manuscript might provide an explanation to the change of exponents close to failure.
Some of the assumptions considered in the approximation to the thermodynamic limit fail at the yield point, where macroscopic effects appear. 
In the standard DFBM this macroscopic effect is limited to a single brittle event. In the GVE-DFBM, the transient hardening at the failure point generates a whole triggering tree with specific statistical properties. As mentioned in section \ref{sec:BD:foreshocks}, these events cannot be expressed as \eqref{eq:BD:aval2} and, thus, are outside the UC. The identification of such non-UC events as post-peak relaxation  might not be possible in finite range interacting systems, where the failure point can be smeared in local interconnected regions due to the material heterogeneity.
\\

This model can set a framework for future experimental studies relating statistical features such as critical failure, ASR and temporal correlations to driving conditions and internal dynamics. Specifically, the proposed relation between triggering and viscoelasticity can be tested in heterogeneous materials with well parametrized viscoelasticity at the microscale by comparing the triggering rates and criticality with the predictions of the MF model or modifications with complex short-range interactions. 
Additionally, we have shown in section \ref{sec:BD:B} that in mean field models of fracture the divergence in $d\Delta/dt$ at failure is determined by the evolution of $B$ as a function of the driving which is difficult to control in some AE experiments \cite{Friedman2012,Maas2015}.
Furthermore, it is difficult to clearly discriminate between stick-slip phenomena and microscopic fracture in some AE experiments of fracture \cite{Baro2013,Davidsen2017}. We have shown in section \ref{sec:BD:distro} that, under the same driving, the exponents related to the divergence of $d\Delta/dt$ are different in the MF approximation of both models. 
The possible mixture of both kind of processes in some cases, related to dynamic weakening \cite{Amitrano2012}, and the variations in the effective driving might explain the variability in the exponent determining the divergence of energy at failure observed in AE experiments  \cite{Zhang2006,Wang2008} within the framework of MF theory. This needs to be addressed in future lab experiments.
\\

This work was financially supported by the Natural Sciences and Engineering Research Council of Canada (NSERC). We thank Gui-Qing Zhang for fruitful discussions.

\section*{Appendix: Numerical implementation}

To validate the analytical approximations presented in this work, we implement the simplest GVE-DFBM, with $\alpha=1$, corresponding to the standard Zener element. 
While the values $S_{i}$ at which an avalanche is activated are absolutely determined by the constitutive curve and the $h$ parameters, the time intervals between avalanche depend also on the relaxation of the hardening, given by the $\alpha$ values. The selection of $\alpha=1$ allows a simple implementation since all the history of the process can be simplified.
For $\alpha=1$, the time dependence introduced in the elements $\phi_{s}$ can be factorized as $\phi_{s}(t+dt)=\phi_{s}(t)\exp(-dt/\tau)$. The inter-event times ($\delta t_{j}$) between consecutive fiber breaking ($j,j+1$) can be found analytically by imposing a fixed external field $\sigma$ in eq.~\ref{eq:BD:transient} leading to the expression:

\begin{equation}
    h e^{-\frac{\delta t_{j}}{\tau}} = \frac{\Phi h + \frac{\delta F}{(1-F_{j})(1-F_{j+1})} + \left({ \Phi h - \frac{\delta F}{(1-F_{j})} }\right)\frac{\delta s}{s_{j}}}
    {\Phi + \frac{\delta F}    {(1-F_{j})(1-F_{j+1})}} ,
\end{equation} 
where $\Phi := \sum_{t_{i}<t_{j}} \phi_{i}(t_{j}-t_{i})$. 
Both avalanches and temporal correlations can be obtained from the right-hand term of this equation. When the term is larger than $h$, the associated inter-event time  ($\delta t_{j}$) is negative and the next fiber will break instantaneously within the same avalanche. For values between $0,h$ we can associate a triggering inter-event time ($\delta t_{j}$) between avalanches. No time can be associated for negative values of the right-hand term, meaning that an increase of the external field $\sigma$ is required to activate the next breaking. This last situation corresponds to the definition of avalanches in the standard DFBM and, hence, defines independent clusters in the GVE-DFBM.\\


\begin{thebibliography}{100}
\bibitem{footnote1} Since the values are discrete a uniform unitarian jitter in the x and y axes has been added to better visualize the density of the point clouds.
\end{thebibliography}
\end{document}